# Teaching Telecommunication Standards – Bridging the Gap Between Theory and Practice

Antoni Gelonch, Vuk Marojevic, Ismael Gomez

antoni@tsc.upc.edu, maroje@vt.edu, ismael.gomez@softwareradiosystems.com

*Abstract*—Telecommunication standards have become a reliable mechanism to strengthen collaboration between industry and research institutions to accelerate the evolution of communications systems. Standards are needed to enable cooperation while promoting competition. Within the framework of a standard, the companies involved in the standardization process contribute and agree on appropriate technical specifications to ensure diversity, compatibility and facilitate worldwide commercial deployment and evolution. Those parts of the system that can create competitive advantages are intentionally left open in the specifications. Such specifications are extensive, complex and minimalistic. This makes the telecommunication standards education a difficult endeavor, but it is much demanded by industry and governments to spur economic growth. This paper describes a methodology for teaching wireless communications standards. We define our methodology around six learning stages that assimilate the standardization process and identify key learning objectives for each. Enabled by software-defined radio technology we describe a practical learning environment that facilitates developing many of the needed technical and soft skills without the inherent difficulty and cost associated with radio frequency components and regulation. Using only open-source software and commercial off-the-shelf computers, this environment is portable and can easily be recreated at other educational institutions and adapted to their educational needs and constraints. We discuss our and our students' experiences when employing the proposed methodology to 4$^{th}$ generation (4G) long-term-evolution (LTE) standard education at Barcelona Tech.

## 1. Introduction

Standards are fundamental for the development of products in many technical areas. Standardization tackles real problems and defines the requirements of a technological ecosystem where a diverse set of players can effectively pursue their business objectives. Any company developing a method, process, service, or device in compliance with a standard needs to pass the homologation process, which consists of a series of tests that are defined in the standard. So, this regulated interaction is the cornerstone that holds the ecosystem and allows interactions (compatibility, interoperability) among the stakeholders (manufacturers, service providers, etc.).

Standards have become a catalyst for technological innovation in numerous areas of science and technology because of the way standards are defined, leaving room for innovation and market differentiation [1]. Standards become a tool to coordinate efforts of various stakeholders while preserving competition. Involved companies can take benefits of economies of scale, build or strengthen collaborations, and participate according to their business model and capability.

The potential of standards to spur economy and impact society is apparent more than ever in the increasingly globalized world. Standards developed by the 3$^{rd}$ Generation Partnership Project (3GPP), a consortium of several standard setting organizations (SSOs) that standardizes cellular communications, have led to an estimated global revenue of more than $3.3 Trillion in benefits and more than 11 Million jobs in 2014 [2]. A Billion human users enjoy wireless communications services today and multiple Billions of machines will be connected very soon.

## 1.1 Problem Formulation

Cellular communications are evolving towards the fifth generation (5G). Five revisions of the 4G long-term evolution (LTE) have been released to date, while in parallel, IEEE and other standardization bodies evolved their WLAN or IEEE 802.xx series products, with a different mobile system profile. Many jobs in the wireless communications industry require telecommunication standards education. Implementing or evolving a complex standard such as LTE is challenging for anyone, but can be overwhelming for fresh graduates. The 3GPP specifications are written in an unusual language, are often intricate and refer to other documents, requiring a steep learning curve. The technical reasons for specifying one parameter or technique over another are difficult to understand and oftentimes have historical, political, or economical foundations. Moreover, typical parameter values that can be useful for implementing an algorithm are extremely difficult to find in the specifications. Despite the minimalistic and formal description, standards have been developed with implementation in mind.

Recent graduates are highly motivated and have strong theoretical background in many aspects of telecommunication systems and may have basic familiarity with modern standards. The skills that are needed to implement a standard-compliant communications system are rare to find. Even after completing a PhD in electrical engineering, graduates often lack implementation skills such as advanced programming or understanding the limitations and constraints of real systems. At the university, a student learns how to solve a particular problem, analyze the available solutions and develop alternative approaches. But, until actually implementing an algorithm and facing the practical challenges in terms of complexity and performance, the student does not fully understand the true differences and practical implications of selecting one algorithm for a standard over another. Therefore, standard-specific implementation, compliance and performance assessment should be components of the electrical engineering curriculum.

## 1.2 Proposed Approach and Related Work

Teaching wireless communications standards is a challenging objective. Important efforts are therefore being made by the IEEE Standards Association (http://standards.ieee.org/about/stdsedu/index.html) and others to show the importance of standards and the role that standardization plays for the industry and society. Through the IEEE Standards Education program, IEEE creates and distributes a variety of educational material and actively promotes the integration of standards into academic programs. They understand that standards are a tool that allows transitioning from theoretical, simulation and experimental results to real-world implementations. Standards combine fundamental concepts with system implementation and address conformance, interoperability, operation and management tasks. We argue that the reasons behind the technical choices, their impact on resources and performance versus flexibility tradeoffs are important components of telecommunication standards education. Moreover, project management, teamwork, development of realistic expectations and practical solutions to imminent problems are skills that are demanded by the industry in addition to the domain-specific technical background. We therefore propose a methodology that allows developing such skills.

The combination of lecture-centered educational methodologies [3] with laboratory-centered approaches [4] [5], has been adopted in the engineering curriculum with special emphasis when the *Conceive, Design, Implement and Operate (CDIO)* methodology appeared in the last decade. CDIO defines a structured methodology to translate the expected education outcomes to the curriculum [6] [7] [8]. Whereas lecture-centered education is considered one of the most effective learning methods [3], it is often criticized for not helping students to transform their knowledge into skills. Laboratory work enhances student skills and helps to consolidate the acquired knowledge. Other cognitive techniques that help addressing the development of the much needed skills include the *scaffolding* approach, where the students receive some support from the instructor, who incrementally reduces this support when no longer needed, the *collaborative learning* approach, where the collaborative process gives students the possibility of sharing thoughts and approach a valid solution [9], or the *student-centered learning* approach that provides support that attends the specific student needs [4].

Considering the nature of standards and attending the industry needs, implementation-orientated active learning methods, such as Project-Based Learning (PBL, http://www.bie.org), provide a student-centered learning environment that is appropriate for the purpose. Learning by doing has been a major engineering education breakthrough, inspired by how humans learn, how they develop expertise and what mechanisms they activate when thinking at higher level [10]. PBL also has its drawbacks. Students typically experience difficulties to initiate their project and do not reach the necessary depth when they lack sufficient background knowledge [11]. An interesting proposal to overcome this issue is the spiral step-by-step method [12], where information is grouped into stages and provided sequentially so that students can better focus and develop the necessary background with sufficient depth.

We propose a PBL methodology for teaching telecommunication standards. We apply this methodology for teaching 3GPP LTE beyond the basics by making use of free, open-source software-defined radio (SDR) development tools. The presented methodology was applied to the master course *Wireless Communications* taught at the Castelldefels School of Telecommunications and Aerospace Engineering (EETAC) of Barcelona Tech. Along with the methodology and case study, we describe our experiences and observations while teaching this course. The methodology and SDR framework are portable and allow adapting to different learning environments and learning objectives.

## 2. Enabling Technologies

A wireless communications standard defines the physical and logical components of the system, the processes and performance requirements. The functionalities are split into basic functions which are formally presented in the specifications, only once, following the established document organization and indexing. These functions comprise algorithms, often expressed as one or more mathematical operations or one or more tables, and interact with other functions through well-defined interfaces to provide the desired functionality.

SDR technology and the availability of open-source software libraries for several digital signal processing (DSP) functions allow implementing complete radio systems in a few laboratory sessions. Software libraries exist for implementing wireless communications standards, such as openBTS implementing the global system for mobile communications (GSM) and Amarisoft, OpenAirInterface and srsLTE for LTE. These software libraries help experiencing these systems at low cost.

SDR technology intrinsically supports hands-on learning, facilitating system implementation and practical analysis. We therefore advocate for using SDR tools to implement, validate, and evaluate the performance of a wireless communications standard. SDR development and implementation frameworks, such as the software communications architecture (SCA)—primarily used in military radios [13]—, GNU Radio—primarily used in research and education (http://gnuradio.org)—, and the application layer and operating environment (ALOE)—also used in research and education (http://flexnets.upc.edu)—, have certain features in common with the specifications of wireless communications standards. SDR frameworks use modular programming and support the concatenation of modules and access to external equipment through common interfaces.

ALOE is an open-source SDR framework that is specifically designed for the implementation of modern radio systems [14]. It takes advantage of the regular data flow of DSP chains and provides a limited set of customizable services. The framework abstracts and virtualizes heterogeneous multiprocessor platforms, provides a packet-oriented network with FIFO-based interfaces between processors, and coordinates the real-time execution of the entire system. ALOE dynamically monitors the computing cost for every processing module and allows observing other critical system parameters in real time as well. Figure 1 provides a screenshot of the working environment of ALOE for a specific experiment. A simple processing chain is defined in the upper-left terminal. The .app file (app for application, i.e., the DSP chain of an SDR transmitter or receiver) specifies the modules, their operational parameters, and their interconnections. The lower-left terminal shows the execution status of the application and the system. SDR or other peripherals can be interfaced through specific modules that use the vendors' APIs. Switching from a simulated channel to over-the-air transmission or reception then involves modifying the .app's sink or source module.

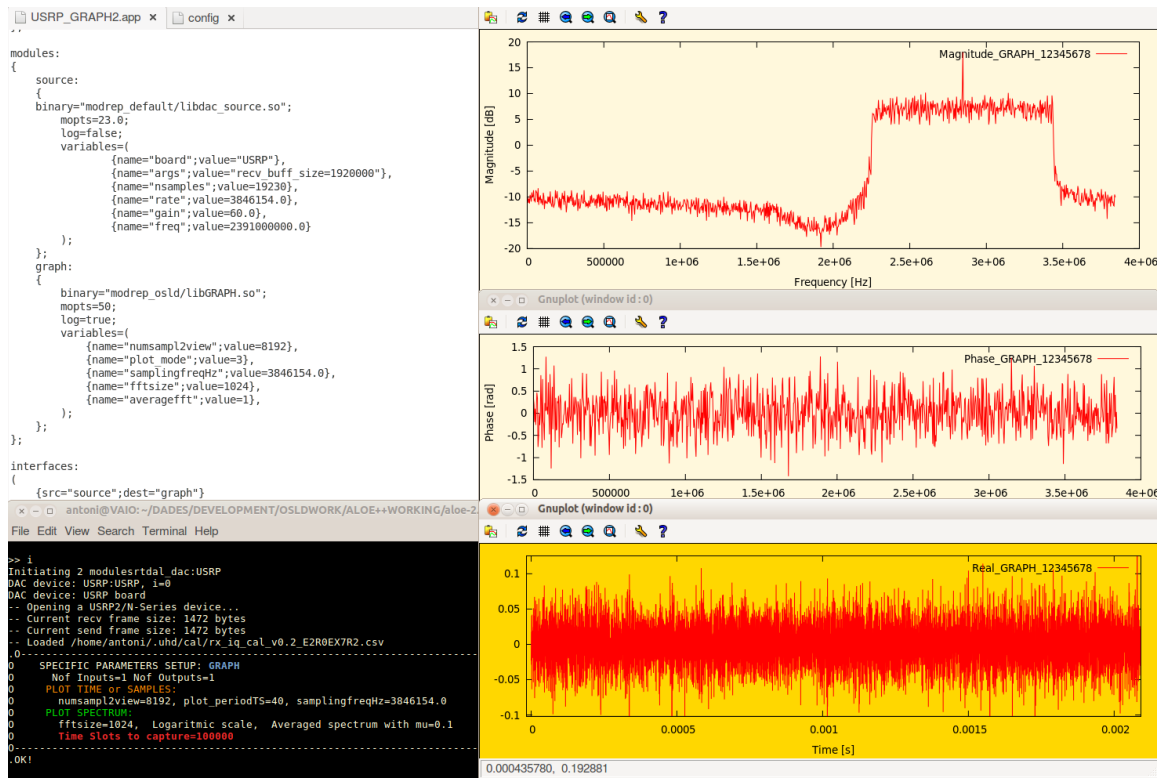

Fig. 1. ALOE working environment. The Upper-left window shows an: .app file which defines the modules involved, their configuration parameters, and the interfaces between them. The lower-left window provides execution control and system status information. The graphs show a 1.4 MHz LTE spectrum and signal. LTE signal was generated and captured using a pair of Universal Software Radio Peripherals (www.ettus.com).

# 3. Teaching Telecommunication Standards: Methodology and Case Study

Instead of reverse-engineering the standard, building the standard out of fundamental building blocks, or functions, aligns with the human learning process. Many basic functions are introduced in prior undergraduate and graduate classes. Here the student can focus on learning how to use these functions in concert and combine them into larger functionalities to achieve the desired system behavior. The assembly of function and the analysis of how these functions work together and how they affect the subsystem or system performance allows gaining invaluable insights into the specifications and reinforces the practice-oriented learning process. Building and testing a prototype that follows the specifications of an industry standard gives significance to the inherent implementation profile of standards and covers many of expected skills.

We provide students with a preliminary system implementation and define assignments that lead to gradually building and testing part of a standardized communications system. The students used ALOE to implement, validate, and evaluate the performance of the adopted solution with respect to its complexity. Students obtain grades from measurable system performance results. The motivation for students to make the system work helps them acquiring a solid background of the technical details of the standard.

Building a prototype that follows an industry standard gives significance to the implementation profile of standards and covers many of the expected skills. Having successfully completed this course, the student will be able to:

- Read a telecommunication standard and find the desired information,
- Design and implement a telecommunication system that it standard-compliant,
- Discuss the pros and cons of alternative technical solutions, and
- Discuss possibly evolutionary paths for the standard being analyzed.

The class is divided in groups of 5 or 6 students. Inspired by the *scaffolding* and *spiral step-by-step* educational methodologies, after providing a high-level overview of the standard under study in the first quarter of the class, we narrow down the focus. More precisely, the students implement part of the system and test its proper functionality, performance, and standard compliance based on previously defined metrics (about two quarters of the class period). Finally, the students discuss the technical decisions that were made during standardization and identify alternative solutions or improvements (last quarter).

The development and testing, being the main part of the class, is continuously monitored by the professor during weekly sessions (2-3 hours), where students describe their progress and the troubles encountered, followed by discussions about the solutions adopted and the progress along the roadmap.

The proposed methodology balances the teaching material and assignments to fit the schedule and accommodate the specific learning objectives that the instructor considers of highest relevance. We propose six learning stages to guide the students through their projects, grouped into the modeling (I), development (II) and evaluation & review (III) phases. These are summarized in Table 1 and discussed in continuation. We provide a brief description of the methodology and exemplify it using a real case study.

Table 1. Proposed learning stages.

| | Learning Stages | Learning Objectives (To be able to…) | Tasks [Instructor] [Students] |
|---|---|---|---|
| I | 1. Overview of the standard | **Good understanding of the standard**<br>• Identify, at a high-level, the critical components of the standard, the relations among key components of the standard as well as some of the important options and tradeoffs<br>• Discuss how and where to search for specific information | **Tutorials**<br>• Standard technology and concepts description<br>• Standardization mechanics and specification documents organization<br>• SDR framework to be used in the project |
| | 2. Abstract modeling | **Design the system**<br>• Assemble a model of the main processing chain of the standard-specific transmitter and receiver<br>• Discuss the processing tradeoffs and how they impact key performance parameters, such as synchronization, throughput, latency, and spectral efficiency | • Propose the project and define the specific assignments and milestones<br>• Develop a complete model of the system<br>• Document |
| II | 3. Narrow the focus | **Define tests and figures of merit (FOMs)**<br>• Identify the key FOM for a system of interest<br>• Design performance and conformance tests based on the FOMs while taking into account the practical circumstances and limitation | • Define Conformance Tests and FOMs<br>• Define Performance Test and FOMs<br>• Document |
| | 4. Development and testing | **Implement and test**<br>• Implement the design from available building blocks<br>• Test the implementation in terms of functionality, compliance with the standard specifications (conformance) and performance | • Provide a baseline implementation<br>• Develop prototype to perform conformance and performance test<br>• Support to validate results<br>• In case of failure propose and perform corrective measures |
| III | 5. Review | **Review the product and process**<br>• Identify where failures happened and discuss short-term remediation techniques as well as long-term solutions<br>• Analyze and design possible system evolution | • Discuss what went right and what went wrong<br>• Document |
| | 6. Publicity and Evaluation | **Demonstrate the product and process**<br>• Demonstrate how objectives have been met and what process has been followed in obtaining the results<br>• Defend the work and discuss alternative approaches<br>• Evaluate the system and the team and individual team member performances | • Demonstration<br>• Poster<br>• Document and software library<br>• Students provide a self-evaluation of the team and individual team members<br>• Instructors evaluate group and individual performances |

The results presented in our case study are extracted from the documentation delivered by the project teams.

This paper discusses the project entitled "Study of the computing cost of the LTE PHY", carried out by multiple student groups in 2012-2014. Starting from a baseline implementation, the project objective was implementing the missing pieces of the physical downlink shared channel (PDSCH), the data channel of LTE, and analyzing the impact of adaptive modulation and coding and the decoder on the system performance, but also on the computing demand.

LTE defines about 30 modulation and coding schemes (MCS) and employs turbo coding and decoding. The LTE base station, or eNodeB, assigns the mobile terminal, or user equipment (UE), the highest possible MCS according to the channel conditions reported by the UE. Changes in the MCS are notified to the UE receiver as part of the control signaling. The UE decodes the control messages first and accordingly modifies the operational parameters of the receiver processing chain. We provided a simplified LTE PHY processing chain through ALOE [14], which features the eNodeB transmitter and UE receiver and a simulated channel. The students download and install ALOE on their computer and do not needed any additional hardware.

### 3.1 General Overview of the Standard

*3.1.1 Methodology*—The student needs to get familiar with the standard and the standardization mechanics. We therefore provide

   a) A high-level description of the standard, from a general description to some details, describing theoretical concepts and employed technologies, identifying relevant working parameters and expected behaviors, and
   b) An overview of the standard specifications and the relationship among the main and auxiliary documents.

According to our working approach and temporal restrictions, we suggest providing tutorials in no more than two or three lecture periods. These tutorials should also cover the SDR framework or tools that the students will use in continuation of the course.

*3.1.2 Case Study*—The instructor provides LTE tutorials that cover the following topics:

- Overall LTE architecture description and functional split,
- Radio protocol architecture: A description of functionalities of user plane and control plane signaling,
- Fundamental resources, timing, multiuser access and scheduling,
- LTE PHY: Logical and physical channels and mapping to physical resources, synchronization process, retransmission protocol, and so forth,
- System performance metrics,
- Conformance test and RF regulation, and
- Organization of LTE specifications with focus on PHY.

The LTE tutorial includes a description how LTE specifications are organized with emphasis on how the Technical Specifications Group Radio Access Networks (TSG RAN) and their working groups (WG) specify the LTE air interface. A flavor of the information provided to the students is shown in Figure 1.

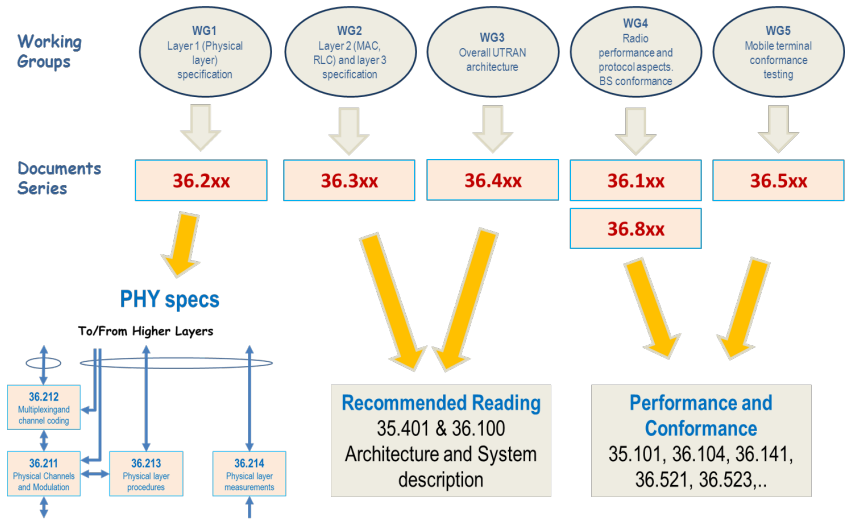

**Figure 1.** Working groups and documents specifying the LTE PHY.

Observing the student progress over the years we found that the tutorials should be defined around a handful of key themes and involve the students. A technique that has worked well is having the students summarize each session based on a template and specific questions that emphasize the key take-home messages. This way the students obtain a general overview of the LTE standard, how the specifications are organized and how to search for details.

Following the LTE tutorials, we introduced ALOE and the tools for building an LTE system. The ALOE tutorial describes the ALOE architecture, tools and services and makes reference to the ALOE Web Site (http://flexnets.upc.edu), where the entire ALOE code base resides and can be downloaded, installed and modified for free.

### 3.2 Abstract Modeling: Modeling the Processing Chain

*3.2.1 Methodology*—For wireless communications standards, the physical layer (PHY) is a key component of the system and is, therefore, a candidate for more detailed analysis. By abstracting the PHY, other parts of the standard can be analyzed instead.

According to the project specification, defined by the professor, the student teams are tasked to develop a system model. This model should identify not only the functionalities (boxes, modules) and their interconnections, but also the working parameters as well as an estimation of complexity. This stage is part of Phase I, where the students develop a model based on the standards overview and available tools.

*3.2.2 Case Study*—Student teams develop a connected graph that illustrates the LTE PHY. One realization is shown in Fig. 2 and illustrates the simplified LTE PHY processing chain of the downlink transmitter and receiver. The colored blocks represent processing functions or processing chains and are specific to the LTE standard (http://www.3gpp.org/). For example, the resource demapping module, RESDEMAP, extracts the control and data symbols and demultiplexes it to be processed by different processing chains. The tables identify the amount of data flowing through the interfaces between the modules for two MCS instances.

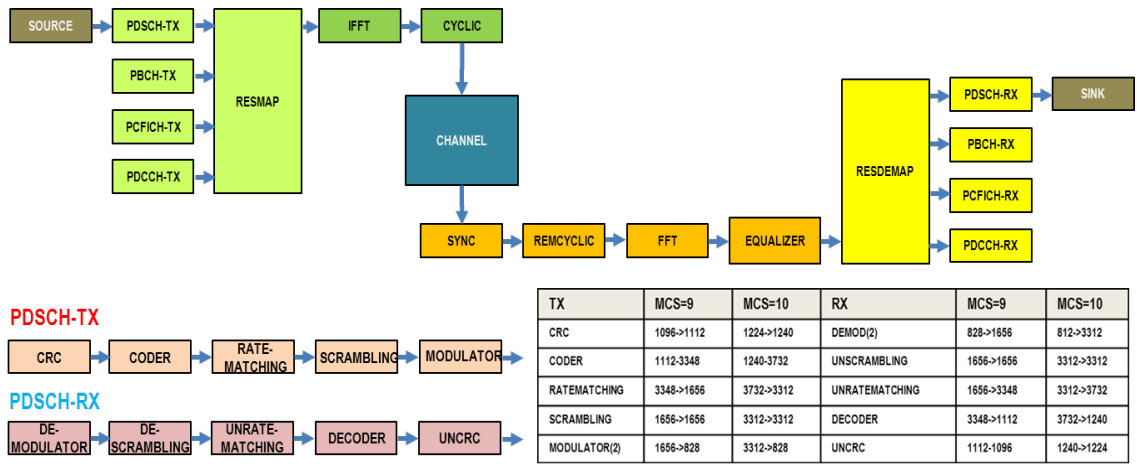

**Figure 2.** Modeling the LTE PHY processing chain.

Such high-level modeling along with the analysis of relations among modules and functionalities and the impact of some of the important parameters provide a good perspective for addressing the partial implementation and analysis of the LTE system.

### 3.3 Narrowing the Focus

*3.3.1 Methodology*—The extension of modern wireless standard specifications and the limited course duration require further narrowing down the focus of the project to specific aspects of the standard. The focus could, for example, be on breakthrough technological concepts that distinguish this standard from its predecessors or emerging concepts incorporated as part of the evolution of a standard. According to the specific project goals, students need to identify those parts of the standard's specifications that require a deeper analysis.

Conformance tests are an important part of telecommunication standards. Performance tests help to understand the system behavior and to identify key figures of merit (FOM). The focus could therefore be to identify standard-specific tests by the students under the close supervision of the instructor:

a) *Define Conformance Tests* to check the suitability of the proposed implementation and fulfill the project specifications based on those defined in the standard.
b) *Define Performance Tests* to address the impact of the employed technologies on the overall system performance.

As a result of both activities, the team develops a set of FOMs to quantitatively characterize the performance of the system w.r.t. the project requirements.

*3.3.2 Case Study*—The objective of the chosen project was to analyze the impact of the MSC on the LTE system performance. By measuring the computing cost, the system performance can be plotted versus computing overhead to emphasize the growing importance of computing in modern wireless systems. The student team working on the project defined the following tests to validate the system and analyze its behavior:

- *Conformance tests:* The first test validates the behavior in terms of bit error rate (BER) of the downlink processing chain when using the three LTE modulation formats that map 2, 4 or 6 bits to modulation symbols and the simulated additive white Gaussian noise (AWGN) channel. The second test checks that the block error rate (BLER) is always below 0.1 or 10 %, according

- *Performance tests:* The performance tests measure the BLER and the computing complexity (processing time overhead) for a selected set of MCS values and different signal-to-noise ratios (SNR) in a simulated channel.

Student team understood how to validate the processing chain according LTE standard specifications for later analyzing system performance.

## 3.4 Development and Testing

*3.4.1 Methodology*—This is the core part of the proposed methodology, where the students actually implement part of the standard they have previously examined (1) and designed (2) and analyze their implementation based on the FOMs (3). The availability of a partial system implementation facilitates this phase and narrows it down to fit the course schedule. Students build the processing chain for their project from the standard specifications using the provided tools and the provided baseline implementation. The system components and the subsystems are continuously validated for correct functionality using test vectors and known output statistics.

The second testing phase evaluates the system or subsystem for conformance, based on the FOM defined in the previous stage. The results obtained from the conformance tests are validated by the instructor. In case of failure, an analysis of the implication on the overall subsystem performance follows. The team then makes a decision whether to continue or solve the problem.

Once the conformance tests are satisfactory, the third testing phase can be initiated. The performance tests are performed and the results analyzed by the team in two sub-stages: (1) analyze the system or subsystem performance w.r.t. the expected performance and discuss the differences, if any, and (2) devise corrective strategies if system performance does not match the expected results or discuss alternative solutions to improve the performance furthermore.

In some cases, both conformance and performance test may require the use of simulated channels, e.g. simulated fading channels, whereas in other cases controlled over-the-air transmission and reception would be more appropriate.

*3.4.2 Case Study*— After having defined the FOMs in the previous stage, the students develop the partial system using a baseline implementation and perform the conformance and performance tests. The following figures and discussion, extracted from the project team's documentation, provides insights about the quality of the work as an indicator of success of proposed methodology.

*A) MCS and System Performance*
Figure 3 plots BLER over MCS for different SNRs with the objective to check the compliance of available implementation with LTE standard specs. The project team realized that demodulation and decoding process requires a certain SNR to achieve the 0.1 BLER target (3GPP LTE specs) which varies according the chosen code rate or MCS.

The students learned how to use the turbo decoder and its relevance for error correction. Whenever the receiver implementation did not fulfill the 3GPP LTE specifications, the number of iterations was increased, from 1 decoding iteration (solid lines) to 5 (round points), in this case (Fig. 3).

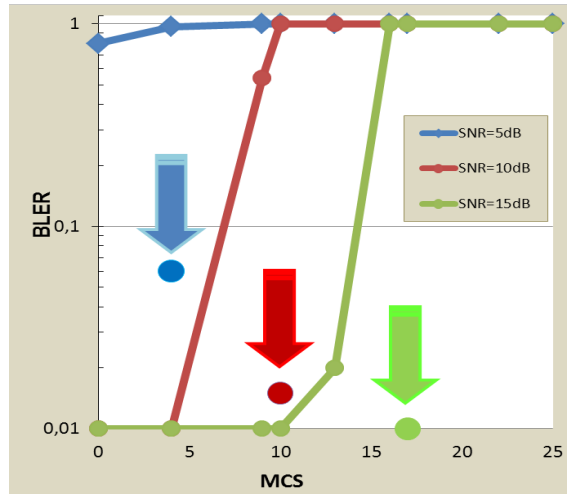

**Figure 3.** BLER versus MCS for different SNR values. The solid lines indicate the achieved BLER with one turbo decoder iteration, whereas the round points correspond to the BLER achieved with 5 turbo decoder iterations.

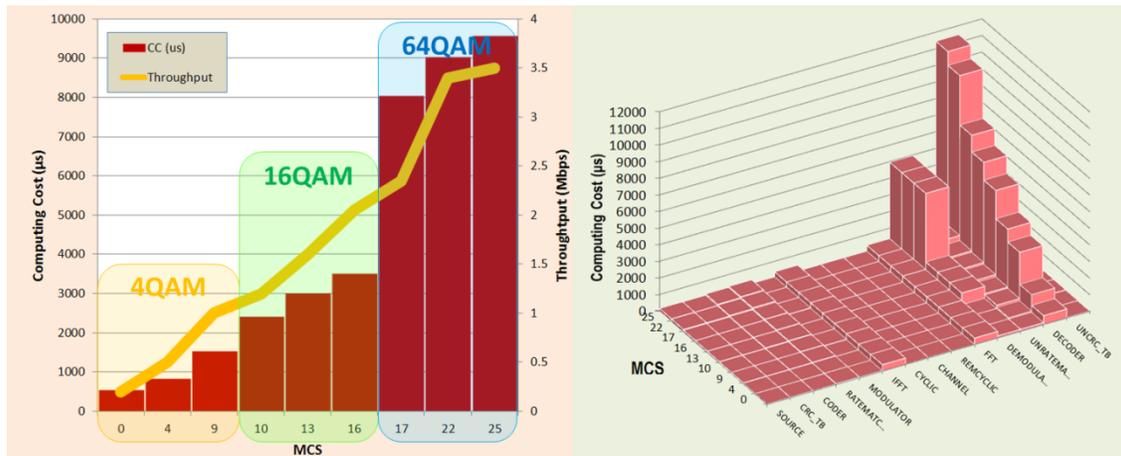

**Figure 4.** Computing cost: Global computing cost and throughput versus MSC for one decoding iteration (a) and computing cost versus MCS of the PDSCH LTE processing chain modules for 5 decoding iterations (b).

*B) MCS and Computing Cost*
Figure 4 plots the user throughput and computing cost for 1.4 MHz LTE and different MCS values. LTE uses three modulation schemes, 4, 16 and 64 quadrature amplitude modulation, mapping 2, 4 and 6 bits to one modulation symbol. The computing cost was a measure of the time spent to execute the processing chain using an ASUS X200CA Netbook PC (Intel Core i3-3217U) and Ubuntu 12.04.3 LTS operating system. Figure 4a results from analyzing relationship between the transport block size—the number of bits transmitted in one transmission time interval—, the number of resource elements and the throughput. The nearly linear relationship between computing cost and MCS matches the expected behavior. According to the 3GPP specifications, an LTE UE can send one of 16 Channel Quality Indicators (CQI) to inform the eNodeB about the highest MCS that it can decode with a BLER not exceeding 10%. Students experienced that more than one MCS can provide the required performance, but each has a different computing cost. Figure 4b shows the computing cost of the main PHY processing blocks that can be found in the PDSCH processing chain of LTE. The students analyze these figures to learn which blocks are critical and need careful (optimized) implementation.

### 3.5 Revision

*3.5.1 Methodology*—After successful completion of the tests, the students discuss what went right and what went wrong. In case of unsatisfactory results, an analysis is conducted to identify the cause. This could need a new design (2), new FOMs (3), a review of the project procedure and goals or even a revision of the standard [15].

*3.5.2 Case Study*—The use of FOMs based on implementation objectives helps to clarify to students team their current status. Starting from provided baseline implementation, students revise their system implementation continuously while progress step-by-step, Corrective measures are taken when misalignments with specifications are detected. Regarding this case study, students experienced how the number of iterations of turbodecoder impact into the BLER but also the computing cost, and discussed about solutions from different point of view.

Along the years, the feedback provided by the students revealed that implementing a wireless standard requires advanced skills and more time. The following list summarizes the student feedback, which helped improving the tools and our methodology over the years:

- An optimized implementation of LTE that meets the timing requirements and FOMs requires experience with code optimization,
- A more detailed documentation of the provided baseline LTE implementation is desired to familiarization with the code,
- Incorporate means to identify potential bottlenecks in the project development early, and
- Unbalanced or uncommitted teams need careful guidance.

### 3.6 Publicity and Evaluation

*3.6.1 Methodology*--The student evaluation is defined at three levels. The first one is based on the delivered documentation that describes the work done, the decisions made and the system performance accomplished. A second level is done through a public presentation and demonstration of the work done to the entire class in a session open to other students and faculty. A third evaluation level is provided from each team member. They, better than anyone else, know the level of commitment and responsibility of each participant in the project team. This approach aims at enhancing the cooperation skills of future engineers.

*3.6.2 Case Study*— Student teams provided a comprehensive document summarizing the standard pieces they have analyzed in more detail, the phases of the project, the realized tests and accomplished results, conclusions, and suggestions for improvement. At the end of each semester, the student teams presented their accomplishments with demos, videos, or posters in a demo/poster session. All class instructors and students assist this session, ask questions and make suggestions. The evaluation is, in part based on how well a group presents its work w.r.t. the class learning objectives and the specific project objectives.

## 4. Conclusions

Testing a prototype or product for performance or standard compliance is a valuable experience for electrical engineering students looking forward to contributing to current and next generation standards. The telecommunication industry is constantly looking for graduates with strong theoretical background as well as hands-on experience. Developing prototypes is always a huge endeavor and dealing with concurrent processes of complex real-time systems is challenging for students. It is difficult to teach these skills as part of the engineering curriculum.

This paper presents a PBL methodology and case study for teaching telecommunication standards. We identify three learning phases— modeling, development, and evaluation & review—, subdivided into a total of six learning stages, and describe our methodology in terms of activities of students and instructors to meet specific learning objectives. Since standards are developed with implementation in mind, using the specifications to build a (simplified) product provides the best way of gaining a solid understanding of the standard. We suggest using SDR technology and the ALOE framework, which provides an effective working environment and baseline implementation for the project development in a confined class period. The staged PBL approach allows identifying the necessary skills, transmitting these to the students and providing an effective learning environment for acquiring them.

We have introduced the methodology into the electrical engineering curriculum at Barcelona Tech several years ago. The students that we had have had different interests and prior experiences. Some were motivated and acted as group leaders. Those students got most out of the class. Other students delivered good work, but too narrow and specific. A balance is needed to gain broad knowledge without abstracting too many details. SDR technology and open-source software frameworks, such as ALOE, provide an ideal framework for experiencing telecommunication standards and learning how to read, implement and analyze the standards specifications in as much detail as considered adequate by the instructor.

## Acknowledgements

This work has been partially supported by the Spanish Government, Ministerio de Ciencia e Innovación, through award number TEC2014-58341-C4-3-R.